# Simulating a Dual-Recycled Gravitational Wave Interferometer with Realistically Imperfect Optics

Brett Bochner[1,2]

## Abstract.

We simulate the performance of a gravitational wave interferometer in the Dual Recycling (DR) configuration, as will be used for systems like Advanced-LIGO. Our grid-based simulation program models complex interferometric detectors with realistic optical deformations (e.g., fine-scale mirror surface roughness). Broadband and Tuned DR are modeled here; the results are also applied qualitatively to Resonant Sideband Extraction (RSE). Several beneficial properties anticipated for DR detectors are investigated: signal response tuning and narrowbanding, power loss reduction, and the reclamation of lost power as useful light for signal detection. It is shown that these benefits would be limited by large scattering losses in large (multi-kilometer) systems. Furthermore, losses may be resonantly enhanced (particularly for RSE), if the interferometer's modal resonance conditions are not well chosen. We therefore make two principal recommendations for DR/RSE interferometers: the DR/RSE cavity must be modally nondegenerate; and fabricated mirror surfaces and coatings must be as smooth as is practically feasible.

1. Department of Physics and Astronomy, Hofstra University, Hempstead, NY 11549, USA
2. Email address: brett_bochner@alum.mit.edu

# 1. Introduction

The initiative to detect gravitational waves (GW's) with large-scale, resonant laser interferometers has begun to enter its active phase, led by international collaborations such as LIGO [1], VIRGO [2], GEO [3], TAMA [4], and ACIGA [5]. Beyond the preliminary goal of just detecting GW's, the ultimate goal is to produce enough high quality GW-data in order to contribute substantially to astrophysical knowledge [6]. To attain the greater sensitivity needed for the latter, much research has been done to plan the eventual conversion of the initial round of detectors to more advanced systems. One important innovation planned for the second generation of large-scale detectors like LIGO and VIRGO (and for the initial generation of mid-scale detectors like GEO-600), is implementing a set of related optical configurations known collectively as "Dual Recycling" (DR).

DR [7] is intended as a system for tailoring a detector's GW-frequency response curves [8], in order to search more deeply for promising signals, while avoiding dominant interferometer noise sources. DR is also is expected to strongly reduce interferometer losses [9], thus lowering the detector shot noise levels, as well as increasing the amount of useful power circulating in the interferometer for GW-detection. These benefits of DR, however, ultimately depend upon how well a real interferometer in the DR configuration will perform when it possesses realistically obtainable optical components. The purpose of this paper is thus to confront the following question: Will a full-scale GW interferometer be able to attain the advanced level of performance expected from DR configurations, given mirrors with realistic deformation levels?

The question of DR interferometer performance has been addressed extensively in the litera-ture, both numerically [e.g., 10-15] and experimentally [9, 16-25], resulting in a generally optimistic outlook for DR. Though such results are encouraging, we must caution that several of the most significant effects of deformed optics in real interferometers may be neglected by these estimations. The experimental work demonstrating the benefits of DR has predominantly been done on small-scale (tabletop or suspended) systems, with arms roughly a hundred times shorter in length than those to be used for the Advanced-LIGO or -VIRGO interferometers; and even the application of DR to GEO-600 [e.g., 26, 27] will not answer all questions about DR performance in the largest systems, which will have multi-km arms and more complex cavity configurations (e.g., Fabry-Perot arm cavities).

Full-scale systems will have dramatic differences from their smaller counterparts. Losses due to beam scattering will be much higher, both because of the long travel paths in the arms, and because of the unusually large beam spot sizes that are necessary. Larger beams sample mirror deformations more extensively -- over bigger areas of the polished mirrors, and over a broader range of significant spatial deformation frequencies [28]. Also, cavity g-factors for large-scale systems will be different, leading to qualitatively different modal resonance and degeneracy behaviors. These differences are magnified by the fact that prototype DR interferometers typically cannot implement the full, multiply-coupled cavity systems to be used by the real advanced-generation detectors. As we will show, all of these distinctly different properties of large-scale systems are important, and may be responsible for degrading DR interferometer performance to a substantial degree unless they are properly dealt with.

Numerical DR work, alternatively, is more easily capable of evaluating the full-scale configurations; but the majority of the numerical work has focused upon the effects of geometrical deformations, e.g., mirror curvature and alignment errors. The more complex optical deformations possessed by real mirrors, however, are of extreme importance, since they are primarily responsi-ble for large-angle scattering losses, and for shifting power into high-order laser modes.

For the study to be presented here, therefore, we use detailed numerical simulations to investigate the performance of a full-scale DR interferometer (similar in design to what will be used for Advanced-LIGO), in the presence of mirrors with "realistically complex" deformations. Our numerical model incorporates a wide variety of optical imperfections, such as mirror surface roughness and substrate inhomogeneities, finite aperture sizes, and losses to due absorption and high-angle scattering. Our tool for this work is a pixelized-grid-based numerical simulation program [29, 30], which has been developed and used for a variety of modeling studies conducted by the LIGO group [e.g., 31-34], and by other collaborating GW groups [35-37]. With this program, we will be able to more realistically estimate how DR performs in optical environments as similar as possible to those of the real advanced detectors.

Regarding the potential implementations of Dual Recycling, we note that one may vary the "tuning" of DR to choose from among several different versions, spanning the range from pure Signal Recycling ("SR" ≡ "Broadband DR") [8], to Resonant Sideband Extraction ("RSE") [38], or anywhere in between ("Tuned DR", "Detuned RSE"). In this paper, we focus primarily upon Broadband and Tuned DR; but the results will often be applicable to RSE as well, and we will extend them qualitatively to RSE where possible.

As will be shown, our results generally echo the optimistic results of prior research, but with several cautionary lessons. First, we find that the beneficial effects of DR will be sharply limited by the quality of the optics, mainly due to strong concerns about the scattering losses and modal degradation caused by mirror surface roughness. Also, one must be extremely careful in choosing the cavity parameters and resonance conditions of the DR interferometer, or incorporating DR can make interferometer performance significantly worse, instead of better, due to the inadvertent res-onant enhancement (instead of suppression) of interferometer loss modes. But the overall conclusion of this paper remains positive: Dual Recycling can indeed be used to obtain a significant improvement in peak signal response and overall interferometer performance. Furthermore, we claim that the results discussed below provide a renewed incentive for producing the best mirrors that are reasonably possible. For example, in the ongoing task of providing specifications for Advanced-LIGO mirrors [e.g., 32, 39, 40, 40.5], a great deal of emphasis is placed (correctly) upon reducing mirror thermal noise; but the surface polish specifications are no stricter for Advanced-LIGO mirrors than for the Initial-LIGO optics, other than being prescribed over a larger mirror surface area. We would suggest that obtaining further reductions in mirror surface deformation amplitudes -- though more difficult for future sapphire optics than for the Initial-LIGO, fused-silica mirrors [32] -- should be a similarly important, parallel goal for optics development.

The ensuing discussion is organized as follows: in Section 2, we provide a brief overview of several important properties of DR, including the predicted GW-frequency response curves for different DR tunings, and DR's theoretical loss reduction capabilities. In Sec. 3, we describe the simulation program that was used for this study, the essential physics of the model, and the basic physical specifications of the modeled system. In Sec. 4, we present the main results from our studies of two principle DR configurations -- Broadband DR and Tuned DR -- describing their successful (and sometimes less than successful) performances in the presence of realistic optics. In Sec. 5, we address ways in which the performance of DR interferometers might (with great care) be improved, including increasing the mirror aperture sizes, and breaking the modal degeneracy of the signal recycling cavity. In Sec. 6, we conclude with a summary of how the results presented in this paper could affect the basic outlook for Dual Recycling as a tool for advanced GW interferometers.

## 2. Theory of Dual Recycling: Its Properties and Benefits

### 2.1. Tailoring the GW-Signal Response Function

Figure 1 depicts the core optical configuration of a large-scale (e.g., LIGO, VIRGO) GW interferometer, augmented here to incorporate Dual Recycling. The basic system is straightforward: the beamsplitter and the highly-reflective end mirrors ($R_4$, $R_5$) form a Michelson interferometer, which converts the differential arm length changes caused by a GW into oscillating output signal fields that are directed through the exit port of the beamsplitter. The partially-transmitting input mirrors ($T_2$, $T_3$) are coordinated with end mirrors $R_4$, $R_5$, to form Fabry-Perot (FP) arm cavities in the "inline" and "offline" arms. These FP arm cavities represent an additional resonant stage, which greatly increases the GW-sensitivity at low-to-moderate frequencies, while imposing a roll-off proportional to GW-frequency above a given frequency determined by the arm cavity storage time (e.g., 91 Hz for Initial LIGO, and significantly lower for Advanced-LIGO [41]).

Other than losses due to imperfect contrast, the carrier laser beam is held to (or, for DC-offset locking [41], *near* to) a dark-fringe at the beamsplitter exit port. The bulk of the carrier power therefore emerges from the "bright port" of the beamsplitter, and it is recovered by the "power recycling mirror" ($R_1$), which creates a "Power Recycling Cavity" (PRC) consisting of mirrors $R_1$, $R_2$, and $R_3$. The PRC increases the stored energy in the interferometer, providing a broadband sensitivity gain [42].

The GW-induced signal fields (at $\nu = \nu_{Carr} \pm \nu_{GW}$), on the other hand, emerge through the beamsplitter's exit port ("dark port"), at which point they may either be allowed to leave the inter-ferometer immediately (for GW-detection), or may be reflected back via the use of a "signal recycling/extraction mirror" ($R_6 \equiv R_{dual}$). The addition of this "Signal Recycling/Extraction Cavity" (SRC/SEC) formed by mirrors $R_2$, $R_3$, and $R_{dual}$, alters the resonant storage time of the GW-induced fields, thus modifying the GW-frequency response of the now "Dual-Recycled" (i.e., power recycling plus signal recycling) interferometer. Depending upon the microscopic tuning of this cavity, the GW-sensitivity peak can either be narrowed around the DC peak response ("Broadband" DR), broadened (RSE), or shifted away from DC (and narrowed) to place the sensitivity peak at some selected value of $\nu_{GW}$ (Tuned DR).

The relevant tunings for this cavity are as follows (*not* considering the effects of the FP arm cavities): Broadband DR uses an *anti-resonant* (for $\nu = \nu_{Carr}$) "SRC", RSE uses a *resonant* (for $\nu = \nu_{Carr}$) "SEC", and Tuned DR uses an *anti-resonant* (for $\nu = \nu_{Carr} \pm \nu_{GW}$) SRC. We note, however, that the carrier beam will be resonant in the FP arm cavities, thus picking up a phase of $\pi$ upon reflection from them [43]. Therefore, for dark-port fields near the carrier frequency (e.g., imperfect-contrast carrier light, or GW-signals at small $\nu_{GW}$), their resonant behavior in the SRC/SEC in the presence of the FP arm cavities will be *reversed* from what is stated above: such fields will be highly *resonant* (with longer storage times, compared to the system with no SRC) in Broadband DR; and for RSE, adding the SEC will *detract* from the coupled-cavity resonance, thus greatly reducing the storage time of such fields in the system, and spreading out the resonant response peak.

The different forms of DR have their varied uses, and their drawbacks. Broadband DR (somewhat a misnomer) has its GW-sensitivity curve peaked at $\nu_{GW} = 0$, so that large values of $R_{dual}$ would narrowband the response too much, restricting sensitivity to a low-frequency range domi-nated by interferometer seismic and thermal noise [43]; but small values of $R_{\text{dual}}$ could be used, more in order to reduce losses (see Sec. 4.1) than to alter the GW response curve. In practice, Broadband DR will probably not be used much for these advanced detector systems [e.g., 19]; but Tuned DR, with its shifted sensitivity peak, could be employed with values of $R_{\text{dual}}$ much closer to unity, in order to achieve strong narrowbanding around selected $\nu_{\text{GW}}$ values of interest. This could be useful in deep searches [44] for particular GW-sources (e.g., non-axisymmetric pulsars [43]), as well as moving the peak away from dominant low-frequency noise sources.

RSE, on the other hand, is primarily used to *flatten* the response peak. In advanced GW-detectors, such RSE-induced peak spreading allows one to use FP arm cavities with very large resonant gains (and thus long storage times), without severely narrowbanding the GW-signal response function of the interferometer. In turn, these high FP-arm gains permit the use of a lower gain for the PRC, thus reducing the incident power (and thermally-induced distortions) on the beamsplitter and other mirror substrates in the PRC. These considerations are very important in the design of future GW-interferometers like Advanced-LIGO [32, 41], and we discuss the practical implications of our work for RSE in Sections 4 and 5.

Sample GW-frequency response curves for Tuned DR are demonstrated in Figures 2 and 3 (with GW-strain amplitude *h* set to 1 here, for convenience). They have been computed analytically, using typical interferometer parameter values (not unlike those in Table I, to be given in Sec. 3), via our program "dual_recyc_IFO_V-M_GW-signal_simulator.f" [29], which is based upon the calculations of Vinet *et al.* [45]. In Fig. 2, the SRC is tuned to optimize the signal response at $\nu_{\text{GW}}^{\text{optim.}} = 500$ Hz, while $R_{\text{dual}}$ is stepped up in stages, from zero (i.e., the "Initial-LIGO" case) to 0.99. In Fig. 3, $R_{\text{dual}}$ is held fixed at 0.9, while $\nu_{\text{GW}}^{\text{optim.}}$ is stepped up in frequency from zero to 900

Hz. Note the important distinction between $\nu_{GW}^{optim.}$, and the actual peak sensitivity frequency of the interferometer, $f_{peak}$: though the SRC is tuned to *maximize* the signal (for one of the two GW-induced sidebands) at $\nu_{GW}^{optim.}$, $f_{peak}$ remains at a lower frequency, primarily because of the conflicting resonance conditions of the SRC and FP arm cavities for interferometers tuned far away from $\nu_{GW}^{optim.} = 0$. Thus $f_{peak}$ does not approach $\nu_{GW}^{optim.}$ until $R_{dual}$ is close to unity (unless one uses "de-tuned" FP-arm cavities [45]). Also, as can be seen in Fig. 3, the "narrowbanded" sensitivity peak quickly flattens out for increasing $\nu_{GW}^{optim.}$, given a fixed value of $R_{dual}$. This is also due to the presence of the FP arm cavities, since signals at larger $\nu_{GW}$ (regardless of the SRC tuning) will be increasingly off-resonance in the FP-arms. These effects -- which will be relevant for GW detectors with FP arm cavities, like LIGO (but not for interferometers without them, like GEO-600 [e.g., 19, 21]) -- would dilute the efficacy of Tuned DR, unless an extremely reflective signal recycling mirror can be used. We will study the feasibility of using large values of $R_{dual}$ in Sec. 4.

**2.2. Loss and Noise Reduction Properties of Dual Recycling, and "Wavefront Healing"**

Besides recycling the GW-induced signal fields (to tailor the response function), the presence of a signal recycling mirror also prevents carrier-frequency "loss fields" at the beamsplitter exit port -- which exist due to imperfect fringe contrast -- from escaping the interferometer immediately, as well. In theory, a highly-reflective signal recycling mirror could sharply reduce the emergence of loss fields at the signal port of the interferometer; and insofar as these loss fields contribute to photon shot noise, DR has the potential to reduce the amount of shot noise competing with the GW-signals, as an added benefit of loss reduction.

Meers *et al.* [7, 9] claimed that a signal recycling mirror could reduce exit-port power losses by a factor roughly comparable to its (power) transmission value, $T_{\text{dual}}$; and power loss reductions of a reasonable degree have indeed been demonstrated in prototype DR experiments [e.g., 18, 21]. We note, however, that this is only true when the leaking exit-port power is composed of modes that are significantly off-resonance in the SRC/FP-arm coupled-cavity system. If some loss modes are resonant (or nearly resonant) in the combined system, then the losses in those modes will be *resonantly enhanced* by DR, rather than suppressed [e.g., 19].

It is well known [e.g., 9, 12, 14] that DR may increase power losses; for Broadband DR, in particular, any principle-mode (i.e., Hermite-Gaussian $TEM_{00}$ mode [46]) power leaking out of the beamsplitter exit port would be resonant in the SRC (cf. Sec. 2.1), as well as in the FP-arms, with this double resonance greatly amplifying the $TEM_{00}$ losses. But since most of the exit-port power losses should be (for small-amplitude deformations) in higher-order, non-$TEM_{00}$ modes [e.g., 9], Broadband DR with small values of $R_{\text{dual}}$ (i.e., the most practical form of Broadband DR) would still be feasible. More importantly, however, in Sec's. 4 and 5 we discuss the very serious problem that exists for Tuned DR, in which all non-$TEM_{00}$ mode losses not resonant in the FP-arms *would* be resonant in the SRC, thus strongly amplifying them in the case of a high-gain SRC with $R_{\text{dual}}$ close to unity.

In cases where the SRC does manage to reduce carrier power losses, though, there is another beneficial effect which may occur: the beamsplitter exit-port power that is recycled by $R_{\text{dual}}$ may be able to re-integrate itself into the interferometer as *useful power*, by converting itself from irregular "loss modes", back into the fundamental $TEM_{00}$ mode from which the GW-signal is generated.

This reclamation of useful power[3] is known as "Wavefront Healing" (or "Mode Healing") [e.g., 14]. It has been anticipated as an important property of full-scale DR interferometers [9, 12], since it would allow them to maintain higher circulating power levels in the presence of imperfect optics than would be possible for an equivalent system without signal recycling. Wavefront Healing has presumably had a beneficial effect for recent DR prototype experiments that have achieved significant reductions in exit-port power losses [e.g., 18, 21].

The success of Wavefront Healing relies on the assumption that the conversion of loss-mode power back into the $TEM_{00}$ mode (via interactions with imperfect optics) will be highly efficient, generally much more efficient than the original conversion process of $TEM_{00}$ power into loss modes[4]. Though counterintuitive, this process works because the multiple resonances of $TEM_{00}$ light in the interferometer (particularly in the modally nondegenerate FP-arms), and the length control systems that hold the coupled-cavities to these resonances, act to drive all circulating power into the $TEM_{00}$ mode. The higher-order modes, being resonantly suppressed in the interferometer -- as well as being blocked from exiting by the two recycling mirrors, $R_1$ and $R_{dual}$ -- may have nowhere else to go, other than turning back into $TEM_{00}$ light [29].

Still, there remains two important caveats to Wavefront Healing. First, as noted above, the SRC resonance conditions for important loss modes must be carefully chosen to resonantly suppress losses, and not to enhance them; otherwise, DR will harm the $TEM_{00}$ power buildup, instead of helping it. And second, potentially lost power can only be reclaimed (and "healed") by the signal recycling mirror if it has not been irretrievably lost from the system before reaching the exit-port, such as would happen for power scattered at high angles, beyond the mirror apertures. These issues will be considered in detail in Sec's. 4 and 5.

---

3. The full Wavefront Healing effect should not be confused with the mere reduction of exit-port power losses; the latter has also been called "mode-healing" [26], or alternatively, the "mode cleaning" effect [18].

4. This re-conversion of non-$TEM_{00}$ power into $TEM_{00}$ power is typically necessary for Wavefront Healing to occur in systems with FP arm cavities; but it is not universally necessary, as we discuss elsewhere [29].

Given the importance of DR as a configuration for advanced GW-interferometers, we address the following questions in this paper: (i) Can significantly tuned, narrowbanded frequency responses with sharp peaks be achieved by a DR interferometer possessing realistically imperfect optics? (ii) Is DR really more tolerant of mirror imperfections than the Initial-LIGO configuration? (iii) Does "Wavefront Healing" exist, as predicted -- and how significant is it for large-scale interferometers? The answers to these questions, as shown below, are somewhat mixed: the desired sensitivity curves of Dual Recycling are indeed obtainable; but increased tolerance to mirror imperfections (and Wavefront Healing) are often less pronounced than hoped for, and are only achieved when one is very careful about the design of interferometer optical parameters.

## 3. The Simulation Program and our Modeled System

As described in Sec. 1, we use a detailed, grid-based modeling program (full technical details given elsewhere [29, 30]) to simulate the performance of DR interferometers in the presence of complex optical deformations. Summarizing here, we note that the program stores spatial information for mirrors and electric field "slices" (at various interferometer locations) on square, two-dimensional grids; the grids are generally 35 cm or 70 cm on a side, and contain perhaps 256 x 256 pixels, for typical runs. These grids can model the fine-scale structure of beam wavefronts and important mirror deformations (substrate inhomogeneities, surface roughness, etc.).

The particular mirror deformation maps used for the simulation runs discussed in this paper have been derived from two measurements of real optical components: a reflection phase map of the polished surface of the "Calflat" reference flat mirror used by the AXAF program [e.g., 47], obtained by LIGO from Hughes-Danbury Optical Systems; and a transmission phase map of a trial LIGO mirror substrate obtained from Corning. These two maps were converted into arrays of (respectively) mirror *surface* and *substrate* deformation maps for use on all of the simulated mirrors, via a process designed to create mirror maps with the same statistical properties as these real-mirror measurements [29, 30, 48]. The substrate maps possess RMS deformations of ~1.2 nm when sampled over their central 8 cm diameters. The original family of surface maps possesses RMS deformations of ~0.6 nm over their central 8 cm diameters, and we refer to this as the "$\lambda/1800$" family of surfaces (with $\lambda \equiv \lambda_{Nd:YAG} = 1.064 \mu m$).

We have multiplied the surface deformation maps by simple scale factors, creating families of more highly-deformed surfaces (labeled $\lambda/1200$, $\lambda/800$, and $\lambda/400$, respectively), in order to evaluate how interferometer performance changes with varying mirror surface quality. We note that such mirror deformation levels are representative of real LIGO mirrors: the currently installed fused-silica mirrors for Initial-LIGO have RMS deformation levels somewhat better than $\lambda/1200$ (i.e., all mirror surfaces satisfy specifications equivalent to ~$\lambda/1350$) [32]; and the preliminary stored arm powers and power recycling gain obtained with those mirrors in the Hanford 2 km interferometer [32] represent a performance roughly equivalent to the best achievable performance using mirrors with RMS variations of around $\lambda/600$ or so, in the full (4km) Initial-LIGO interferometers [29]. The mirror map families to be simulated here thus represent an optical quality range of interest for contemporary GW-interferometer systems.

In the simulation program, each short-distance interaction of an electric field with a nearby mirror is done via the ("near-field") pixel-by-pixel multiplication of the electric field grid map with the mirror map [29, 46]. But longer propagations of fields over macroscopic distances are done via Siegman's method, utilizing the paraxial approximation [46], and are carried out via Fourier transform methods using FFT's [e.g., 49]; hence our simulation program is colloquially referred to in LIGO as the "FFT Program" [e.g., 33].

To fully solve for the resonant, steady-state behavior of the fields in the interferometer, the program employs a number of routines for accelerated field relaxation [29, 30, 50] and interferometer sensitivity optimization [29, 30]. The latter refers to optimization steps such as: performing microscopic cavity length adjustments (in the PRC, SRC, and FP arm cavities) to achieve the proper resonance conditions for the carrier and GW-induced signal fields; and, finding the value of $R_1$ that maximizes the interferometer power buildup, given the losses of the optically imperfect system.

To calculate the GW-signals that would be produced by the interferometric detector, the FFT program considers the sidebands (at $\nu = \nu_{carr} \pm \nu_{GW}$) that are impressed upon the FP arm cavity light by a GW with frequency $\nu_{GW}$ and strain amplitude $h$ [29, 30, 45]. The program explicitly simulates the resonant buildup of these GW-sidebands, after their initial generation in the FP-arms, during their "free propagations" throughout the interferometer (i.e., neglecting the continuing but negligible GW effects upon them [29]). The initial data input to the FFT program includes a discrete list of GW-frequencies, and the GW-sidebands for each of these frequencies are modeled individuality, with the overall state of the interferometer (other than the resonance tuning of the SRC) having been fixed by an initial simulation run done for the carrier field. For each frequency, the (single-sided) GW-response is computed by summing together the field amplitudes for plus- and minus-sidebands at the SRC output/detection port. Though this simple summation neglects the complicated process of optimally extracting the signal information from both sidebands simultaneously [e.g., 17], it is a decent approximation at this level of analysis. One can then plot the simu-lated GW-frequency response by connecting the data points over the modeled frequency range, as will be done in Sec. 4.

We note that this procedure only computes relative signal amplitudes, not absolute GW-signal strengths; the latter would require us to assume and model a specific GW-detection scheme for DR interferometers. Although we have used the FFT program to model the control system for the Initial-LIGO detector [29, 30], we have not yet explicitly modeled any of the (complex and still-evolving) detection schemes being considered for Advanced-LIGO, and for other DR interferometers [e.g., 21, 24, 27, 41, 51-53]. Nevertheless, the relative GW-response curves generated from these simulations are very informative about essential features of DR interferometer performance.

The common optical parameters used for each of the simulation runs to be discussed below are given in Table I. Note that these parameters are essentially those for an Initial-LIGO interferometer, except for the addition of the signal recycling mirror. We have opted to keep the interferometer parameters as similar as possible to Initial-LIGO for the runs to be demonstrated here, in order to facilitate direct comparisons between the performance of "initial" and "advanced" interferometer configurations, given equivalent optics. Though the actual applications of DR would likely come in "enhanced" interferometers possessing upgraded interferometer parameters (e.g., lower mirror losses, higher input laser power, different Fabry-Perot cavity storage times, etc.), our studies here are intended as detailed, proof-of-principle tests of the various claims that have been made about the advantages of DR, for the which use of Initial-LIGO optical parameters produces straightforward answers.

# 4. Results of Dual Recycling Simulations

## 4.1. A Brief look at Broadband Dual Recycling

We begin this section with some results from Broadband DR simulation runs, in order to demonstrate that this configuration behaves as expected in terms of its loss reduction capabilities, and to show that our results are in line with those from prior research.

Figure 4 plots the amount of lost carrier power escaping at the interferometer exit port (i.e., through the SRC), for a series of Broadband DR runs with different values of $R_{\text{dual}}$. Each run uses our deformed substrate maps, plus deformed surfaces from the $\lambda/800$ family of maps (effectively large deformations, causing ~50% reductions to the stored FP-arm carrier power). $R_{\text{dual}}$ is varied here from "*zero*" (i.e., the complete absence of the mirror and all of its effects), to 0.9. Note that a short SRC length (5 m) is used here (and for all of the runs in Sec. 4), thus making the SRC a modally degenerate cavity.

The results in Fig. 4 bear out the standard expectations for Broadband DR [9, 11, 12, 14]: the presence of a signal recycling mirror effectively suppresses the *total* amount of power lost, given that most of the losses here are in non-TEM$_{00}$ modes; but the loss in the TEM$_{00}$ mode itself grows with increasing $R_{\text{dual}}$, because it is resonantly enhanced by the SRC (cf. Sec. 2.2). The imposition of Broadband DR would therefore make the exit-port losses worse for values of $R_{\text{dual}}$ too close to unity, or for optical deformations (e.g., large mirror alignment errors [12]) that cause relatively large amounts of TEM$_{00}$ power to leak into the SRC. Provisionally, however, Broadband DR does manage to reduce interferometer losses for the type (and level) of mirror deformations modeled here, given these values of $R_{\text{dual}}$.

## 4.2. A Tuned DR Study with $\nu_{GW}^{optim.} = 200$ Hz

We now study a system with Tuned DR, which is tuned to GW-frequencies high enough to be above the regime where the interferometer's non-shot noise sources dominate [43], yet not so high (cf. Fig. 3) that the DR sensitivity peak flattens out to uselessness for reasonable values of $R_{dual}$, given the FP-arm cavity finesses assumed here. We also consider the important frequency regimes for astrophysical objects of interest: such as an inspiraling binary Black Hole ("BH/BH") system (with equal-mass BH's weighing, say, 10 $M_{Sol}$ each), beginning its merger phase at $f_{merger} \approx 205$ Hz, and shutting off its GW-emissions at $f_{high} \approx 1430$ Hz [54]; or non-axisymmetric pulsars emitting GW's within a typical range of $f_{pulsars} \approx 20 - 1000$ Hz [55]; or GW-radiation from the cores of supernovae within a similar frequency range [56]. Given these considerations, we evaluate Tuned DR here at two different frequency tunings: first, we will present a comprehensive study of a DR interferometer tuned to $\nu_{GW}^{optim.} = 200$ Hz; later, this will be followed (in Sec. 5.2) by a smaller selection of runs with $\nu_{GW}^{optim.} = 1000$ Hz, in order to study some of the problems (and possible solutions) for Tuned DR at high tuning frequencies.

Each run to be discussed has been done either with "perfect" mirrors (i.e., perfectly smooth surfaces and substrates), or with deformed substrates for all of the mirrors, in addition to a full set of deformed surface maps taken from one of the $\lambda/1800$, $\lambda/1200$, $\lambda/800$, or $\lambda/400$ families. The GW-signal response data points for the runs are computed from FFT program simulation results, as per the discussion in Sec. 3.

Figure 5 shows a series of plots, each depicting the GW-signal response for a different value of $R_{\text{dual}}$, from "zero" to 0.99. Each plot contains the results for 3 different runs: "perfect" mirrors, $\lambda/1800$ surfaces, and $\lambda/800$ surfaces. Shown against them are the theoretical curves (cf. Fig's. 2, 3) computed for mirrors with the same values of $R$ and $T$, except that an additional 2 parts per million of loss have been included (in the theoretical curves) for each of the FP arm cavity back mirrors, to account for the diffraction losses that occur there, due to large beam spot sizes encountering finite-sized mirrors [29].

Several important conclusions can be drawn from these (and subsequent) plots: (i) The "perfect mirrors" runs match the theoretical predictions very well, demonstrating the accuracy of the simulations. (ii) The realistically-deformed mirror maps do not reduce the GW-signal levels by huge factors, as long as one does not use very large deformations and/or $R_{\text{dual}}$ values too close to unity. (iii) The response peaks of Tuned DR remain sharp, and are not excessively broadened by the presence of realistically-deformed mirrors. (This lack of broadening is due to the fact that the losses caused by fine-scale mirror imperfections are primarily due to the scattering of power completely out of the system in the FP-arms, which simply reduces the overall reservoir of power stored there for GW-signal field generation.) Thus the main effect here of deformed mirrors is to reduce each response curve by some moderate, generally broadband factor.

To quantify statements (ii) and (iii), consider Figure 6. Curves are presented for each value of $R_{\text{dual}}$, showing the fractional reduction -- as compared to the "perfect mirrors" case, for *that same* value of $R_{\text{dual}}$ -- of the GW-signal at $f_{\text{GW}} = 200$ Hz, as the mirrors are progressively degraded from the perfect case to the $\lambda/400$ case. Since a large $R_{\text{dual}}$ implies a long storage time for the GW-sidebands in the SRC/FP-arms system, and thus more extensive sampling of the mirror deformations, one would expect the most strongly narrowbanded GW-signal response curves to be most susceptible to the effects of imperfect mirrors[5]; and this expectation is borne out by Fig. 6. We therefore wish to know how close to unity $R_{dual}$ can be brought, without the losses becoming so large that the expected, narrowbanded peak is not achievable with realistically-deformed mirrors.

Fig. 6 demonstrates that runs with $R_{dual}$ as high as 0.9 suffer little more *relative* harm to their GW-signal amplitudes, due to deformed mirrors, than the run without Dual Recycling at all would experience; but going as high as $R_{dual} = 0.99$ causes the system to incur significantly higher losses from severely deformed mirrors. For $\lambda/400$ mirror surfaces, in particular, the $R_{dual} = 0$ case retains ~47% of the GW-signal that it would have had (at 200 Hz) for perfect mirrors, while $R_{dual} = 0.9$ retains fully ~44%; but $R_{dual} = 0.99$ retains just ~30% of the GW-signal that it would have had for perfect mirrors. These results are significant, because of the advantages of using very narrowbanded sensitivity peaks for GW-searches (e.g., peaks with FWHM as small as ~6 Hz for coalescing BH binaries, and ~3 Hz for periodic GW sources [44]). For this SRC tuning and these interferometer parameters, $R_{dual} = 0.99$ yields a sensitivity peak with a FWHM of ~8 Hz, while $R_{dual} = 0.9$ only narrows it to ~50 Hz [29]; and we see here that the long storage times and tight narrowbanding in the former case are not so well achieved with highly-deformed mirrors. The use of very high quality mirrors is therefore of great importance, if one is to achieve a truly narrow-banded GW-response function for Tuned DR.

---

5. Note that this is very different from the "Wavefront Healing" effect described in Sec. 2.2, which is supposed to make the losses *smaller* for large $R_{dual}$, by reducing (and recycling) the *carrier* field power lost due to imperfect beamsplitter contrast. But the issue considered here refers to the losses experienced by the *GW-sideband* fields, which are larger for longer storage times in a SRC with significantly deformed mirrors. This effect works against Wavefront Healing, and causes the increase in losses (with increasing $R_{dual}$) that is apparent in Fig. 6.

In light of the fact that increasing $R_{dual}$ makes the interferometer more susceptible to deformed mirrors -- despite the fact that signal recycling was supposed to make the interferometer *less* vulnerable to deformation-induced losses (cf. Sec. 2.2) -- we must ask the question: "Where was Wavefront Healing?" The answer is that Wavefront Healing was indeed happening, but that the practical effect was very small. Wavefront Healing acts by reducing carrier losses at the exit port of the interferometer; but re-examining Fig. 4 from the Broadband DR runs, for example, we see that although DR did reduce the lost exit-port power, that lost power was always less than ~2.5% of the excitation laser power in all cases. Similarly, Figure 7 depicts the lost power (as a fraction of laser excitation power) for these Tuned DR runs, versus $R_{dual}$ and mirror deformation amplitude. Again we see that DR does succeed in reducing the exit-port power losses[6]; but those losses are always less than ~4% anyway, without DR, even for the worst mirrors modeled. Thus the amount of Wavefront Healing which could theoretically be achieved by recycling the exit-port carrier power is not very large.

This is due to the fact that most of the losses caused by mirrors with "realistic", high-spatial-frequency deformations, are *scattering losses* in the long FP arm cavities. Unlike contrast-defect power considered "lost" at the beamsplitter exit port, "light which is scattered at high angles is gone for good" [9], and cannot be caught by $R_{dual}$ for recycling by the SRC. Consequently, Wavefront Healing can only restore FP-arm power to a very limited degree, when scattering in the FP-arms is a primary source of loss.

---

6. DR does not *always* reduce exit-port losses: as shown in Fig. 7 (and subsequently in Fig. 8), a "little" DR (i.e., $R_{dual} = 0.1$ or $0.3$) can actually be worse than no Dual Recycling at all. This is due to modal degeneracy effects in the SRC, which will be discussed in Sec. 5.2.

For example, in our runs with λ/800 mirrors, we estimate [29] that about half of the laser power dissipated in the system is typically lost via absorption in the mirrors (i.e., the "best" way to lose power, since high absorbed power means high circulating power). But only a few percent of the dissipated power is due to exit-port losses (cf. Fig. 7). Thus the remainder of the lost power -- *nearly half* of the laser power dissipated in the system -- is lost via high-angle scattering in the long FP-arms, dwarfing the amount lost through the interferometer exit port as contrast defect. As a result, it can be shown [29] that signal recycling does manage to increase the stored FP-arm power for these Tuned DR runs, but by less than 4%, at most. Thus the Wavefront Healing process does indeed work for these runs, but at a barely significant level, since the dominant types of loss here (scattering and absorption) cannot be "Healed".

As a counterpoint to these results, it can be shown that Wavefront Healing is very effective (as expected) at restoring lost power for cases with simpler optical deformations, such as mirror tilts. Figure 8 shows the results of a study we performed in which the back mirrors of the two FP-arm cavities were (antisymmetrically) tilted by various misalignment angles. Examining the amount of (almost entirely $TEM_{00}$) power circulating in the inline FP-arm for various tilts and values of $R_{dual}$, it is clear that Wavefront Healing is very good at restoring circulating FP-arm power for cases like this, with mirrors possessing only geometric mirror deformations, rather than realistically complex surface deformations.

This limitation to the Wavefront Healing property of DR in interferometers with realistically-deformed mirrors, though dramatic, has not been readily apparent in prior studies. As discussed in Sec. 1, this is because analytical and numerical research into DR interferometers has primarily concentrated upon geometric mirror deformations, which do not cause large amounts of high-angle scattering; and because no experiment with a tabletop or small-scale suspended interferom-eter would have exhibited the kind of scattering losses that would exist for full-scale interferometers with multi-km arms and large beam spot sizes.

These considerations are especially important for DR configurations tuned to RSE. As noted in Sec. 2.1, the goal of RSE is to shift the burden of high resonant gain away from the PRC, and onto the FP arm cavities; but if the FP-arms experience excessive scattering losses, then this will undercut the ability of the PRC to produce any significant gain at all, since the maximum possible PRC gain is limited by those FP-arm losses. For example, in a typical Initial-LIGO run with $\lambda/800$ mirrors (cf. Table 3.1 of [29]), the maximized PRC power gain is only ~38 (compared to ~72 for "perfect" mirrors); and $\lambda/400$ mirrors reduce the PRC gain further, to ~16. These results are passably better than the Advanced-LIGO *requirement* of (roughly) ~15 [e.g., 32, 41]. But that Advanced-LIGO requirement implicitly assumes the use of FP arm cavities with a much higher finesse, which implies even higher scattering losses in the FP-arms, further lowering the achievable PRC power gain. To make RSE workable, therefore, it is crucial to limit scattering losses in the arms, quite possibly by using better mirrors than would be necessary for non-RSE configurations with lower FP-arm storage times. We conclude that for DR to work as well as anticipated in terms of Wavefront Healing -- and perhaps for it to work at all in the case of RSE -- great attention must be paid to the quality of the optics, particularly to the fine-scale deformations of the mirror surfaces, substrates, and coatings.

Despite these drawbacks, the results shown above nevertheless demonstrate that Tuned DR is capable of reducing the amount of exit-port lost power which does exist, thus reducing the exit-port shot noise, and also enabling a "modest" amount of Wavefront Healing to occur. It is also capable of obtaining a sharp, narrowbanded sensitivity peak for the interferometer, when the mirror deformations are not too large; though we must also caution here that much of these Tuned DR sensitivity peaks would be drowned out by a thermal noise "barrier" [29], unless the level of mirror internal thermal noise can somehow be reduced (such as through the development of sapphire optics [32]). Given all of these considerations, our main conclusion from this section is as follows: the benefits of DR that are required of advanced GW-detector systems, configured either as Tuned DR or RSE, cannot be realized without the use of *extremely good mirrors*.

## 5. Problems and Possible Solutions for Dual Recycling

### 5.1. Scattering losses and large-aperture mirrors

Since we find the main limitation to Wavefront Healing to be large amounts of high-angle scattering loss from deformed mirrors, one possible solution to this problem (as suggested by Meers and Strain [9]) might be the use of large mirrors. Mirrors with larger apertures would be able to "catch" a lot of this scattered power, keeping it in the system for recycling and possible Healing.

To test this idea, we performed sets of runs in which the aperture radii of all interferometer mirrors[7] were varied from 12 cm to 35 cm. These series of runs were done with perfect mirrors and with $\lambda/800$ mirrors, for $R_{dual} = 0$ and for $R_{dual} = 0.9$ (i.e., Tuned DR set to $\nu_{GW}^{optim.} = 200 \text{ Hz}$). Figure 9 shows the (inline) FP arm cavity circulating power for all cases.

First of all, Fig. 9 shows that enlarging the mirrors does initially succeed in capturing much of the scattered light, up to aperture radii of ~20 cm; the circulating arm power increases even for the system with $R_{dual} = 0$, with the DR system doing even better. Furthermore, the power increase in the arm cavity is predominantly in the $TEM_{00}$ mode, affirming the notion of Wavefront Healing.

---

7. The beamsplitter was also circularized and increased in size, to match the other mirrors. Additional test runs were done for the 12 cm aperture runs, using the normally elliptical (i.e., tilted-circle) beamsplitter profile, and will also be included in Figure 9; the exact beamsplitter profile shape is shown to make little difference.

For even larger apertures, however, the arm power level flattens off for the non-DR case -- and it plummets for the DR case. This happens because the power lost through the interferometer exit port (not shown) increases greatly for these runs, and becomes dominated by $TEM_{mn}$ modes for which $m + n = 23$. This is due to an accidental resonance for all of these modes (which are mutually degenerate with one another [46]) in the FP-arms. Normally, most of the power in these high-order modes would fall outside of the mirror apertures, and be eliminated from the system; but enlarging the apertures beyond ~20 cm or so allows the mirrors to contain most of these unwanted modes [29], creating a parasitic resonance that siphons off power from the $TEM_{00}$ mode, and which increases the contrast-defect losses at the interferometer exit-port.

Using larger mirror sizes is therefore a very risky strategy for increasing the amount of useful power in the arm cavities. Unless one can guarantee that no "accidental" resonances will occur -- something nearly impossible to do, since complex mirror deformations will scatter power into a large range of high modes -- trimming the fields with reasonably sized mirrors is essential. Recapturing scattered power is no substitute for reducing scattering in the first place; once again, mirror quality is the key.

But the more serious issue for our purposes, is how badly DR performs for the largest-mirror runs -- significantly worse, in fact, than the case without signal recycling. We explain this by recalling the discussion of resonance conditions in Sec. 2.1, and the plot of $TEM_{00}$ power losses in Fig. 4. For Broadband DR, any nonresonant modes in the FP-arms will be anti-resonant (i.e., suppressed) by the SRC. But any modes which are *resonant* in the arms -- like the $TEM_{00}$ mode, or TEM$_{mn}$ with $m + n = 23$ -- will pick up a phase of $\pi$ in reflection from them, and thus will *also* be resonant in the SRC. This turns the FP-arms and SRC into a doubly-resonant, high-gain system for pumping such modes out of the exit-port. For this particular Tuned DR case with $\nu_{GW}^{optim.} = 200\ Hz$, the tuning was close enough to Broadband DR for this doubly-resonant loss enhancement to occur, as demonstrated by Fig. 9. A related problem[8] was also demonstrated in Fig's. 7 and 8, in which it was apparent that a "little" signal recycling was worse than none; this is because the circulating (non-TEM$_{00}$) modes were semi-resonant for that Tuned DR case, and thus an intermediate value of $R_{dual}$ was most "successful" at pumping losses out of the system. A fully-resonant mode in the SRC, however, would be maximally lossy for very high values of $R_{dual}$.

Thus we see that unless all cavity parameters are chosen carefully, the interferometer may experience what might be called "Wavefront Harming" due to DR, instead of Wavefront Healing. Such behavior was seen in the study of DR with geometric deformations by McClelland *et al.* [12], in which their SRC magnified the losses in the TEM$_{20}$ and TEM$_{02}$ modes generated by mirror curvature mismatch. Their solution to the problem was to change the mirror curvatures, in order to push those modes off-resonance. But such a solution for controlling SRC-magnified losses would be much more challenging for mirrors with realistically-complex deformations, as we study here, in which some amount of power would be channeled into innumerable high-order modes.

---

8. The problem from Fig's. 7 and 8, however, was due to the SRC being tuned somewhat *away* from Broadband DR; this general phenomenon for all non-TEM$_{00}$ modes will be explained in Sec. 5.2.

## 5.2. SRC degeneracy and degeneracy-breaking: The significance for RSE and Tuned DR runs with $\nu_{GW}^{optim.} = 1$ kHz

A very important factor regarding this so-called Wavefront Harming effect, is the tuning of the SRC/SEC cavity. As the SRC resonance is tuned to higher frequencies -- thus moving the system from Broadband DR to Tuned DR, and ultimately to RSE -- the $TEM_{00}$ mode (plus any other modes accidentally resonant in the FP-arms) eventually lose their double resonance, and the exit-port losses in those modes become increasingly suppressed by $R_{dual}$.

Non-$TEM_{00}$ modes, on the other hand, experience the opposite effect: the higher the SRC tuning, the more resonant they become there. Exit-port losses in non-$TEM_{00}$ modes are thus *resonantly amplified* by the SRC for high frequency tunings; and this problem is most severe for RSE, in which virtually every non-$TEM_{00}$ mode would be resonantly pumped out through the SEC.

The source of this problem is the modal degeneracy of the SRC/SEC cavity. In large-scale DR systems with FP arm cavities, like Advanced-LIGO or -VIRGO (but not GEO-600), the FP-arms are separated from the SRC by the FP-arm input mirrors (i.e., $T_2$, $T_3$). This potentially makes the SRC a short cavity with minimal beam focusing, and thus modally degenerate [46]. In that case, all modes will experience the same round-trip phase (other than the $\pi$-phase FP-arm reflection) as the $TEM_{00}$ mode, so that tuning this cavity for RSE (i.e., resonant for $TEM_{00}$) will make *all* modes resonant there; or, for Tuned DR, semi-resonant. Because of this, it is imperative that the degeneracy of the SRC/SEC be broken, so that the proper cavity setting (i.e., Tuned DR or RSE) can be achieved for the $TEM_{00}$ mode, without simultaneously forcing all non-$TEM_{00}$ modes towards resonance, as well.

The SRC degeneracy is broken by increasing the amount of beam focusing that occurs during SRC propagation. This can be done either by using strong-focusing optics, or by making the SRC significantly longer. The former is difficult to model in a grid-based program, because strongly-curved mirrors would radically change the angular distribution of energy in the beam [29]; we have therefore chosen to do the latter in our simulations, implementing a 2 km long SRC (e.g., putting $R_{dual}$ at the LIGO mid-station [51]).

As noted above, SRC degeneracy effects were apparent in Fig's. 7 and 8 -- intermediate values of $R_{dual}$ (i.e., 0.1 and 0.3) caused the worst interferometer performance for those Tuned DR runs with $\nu_{GW}^{optim.} = 200$ Hz, because of the semi-resonant amplification of non-TEM$_{00}$ exit-port losses. We have shown elsewhere [29] that breaking the degeneracy with a 2 km SRC can improve DR performance for that tuning, making the exit-port losses decrease monotonically with increasing $R_{dual}$. Here we present the more dramatic effects that SRC degeneracy-breaking achieves for Tuned DR with $\nu_{GW}^{optim.} = 1$ kHz, a tuning near the high end of the interesting GW-spectrum (cf. Sec. 4.2), and closer to the RSE setting in which non-TEM$_{00}$ losses are maximized.

Three sets of runs (each with $\lambda/800$ surfaces and deformed substrates) are included here. They were done using, respectively: (i) A short (~6 m), degenerate SRC; (ii) A long (2 km), non-degenerate SRC ("Config. I"), for which TEM$_{mn}$ modes with $m + n = 6$ were accidentally close to resonance in the SRC/FP-arms combined system; (iii) A long (2 km), nondegenerate SRC ("Config. II"), for which the cavity g-factors were changed slightly to move all significant loss modes away from resonance.

The results for these three cases are depicted in Figures 10 and 11, which show (respectively) the exit-port power losses, and the (inline) FP arm cavity stored powers, as a function of $R_{dual}$. These plots clearly show that Tuned DR with a degenerate SRC can *greatly increase* the exit-port losses, significantly harming the signal-to-noise ratio of the interferometer. Furthermore, this Wavefront Harming effect is much larger than the anticipated Wavefront Healing effect could have been (~3-4 times larger, in this case), and it persists up to high values of $R_{dual}$ (and would persist for almost all values of $R_{dual}$, for the RSE tuning). On the other hand, the plots also show that these problems can be alleviated -- and a modest amount of Wavefront Healing can be restored -- by breaking the SRC degeneracy, as long as accidental resonances (e.g., Config. I) of significant deformation modes in the interferometer are avoided. Doing so is absolutely necessary for Dual Recycling to operate successfully, especially for tunings far from Broadband DR.

## 6. Conclusions

We summarize our conclusions, for Dual Recycling in advanced GW-detectors, as follows:

(1) GW-response curves with moderate narrowbanding (e.g., $\Delta \nu_{FWHM} = 50$ Hz at $\nu_{GW}^{optim.} = 200$ Hz) are achievable, even with relatively bad (e.g., $\lambda/400$ RMS) mirrors; but the efficient use of tight narrowbanding (e.g., $\Delta \nu_{FWHM} = 8$ Hz, at that tuning) requires significantly smoother mirror figures.

(2) In order to get a substantial benefit from the sharp sensitivity peaks attainable with DR interferometers (even in the presence of realistic mirror deformations), the well-known problem of mirror internal thermal noise must be controlled as well as possible.

(3) The Wavefront Healing effect can succeed at boosting stored interferometer power and reducing exit-port shot noise; but this effect will be much smaller than expected for large-baseline GW-interferometers, unless scattering losses in the FP arm cavities can be limited by reducing the levels of fine-scale deformations in the mirrors.

(4) Increasing the sizes of the mirror apertures is a risky method for reducing FP arm cavity scattering losses, since it may backfire by greatly increasing the interferometer's exit port losses.

(5) Rather than reducing exit-port losses (i.e., Wavefront Healing), the use of signal recycling can substantially *increase* those losses ("Wavefront Harming"), if any (or all) of the important loss modes happen to be resonant in the SRC. Furthermore, Wavefront Harming can be a significantly larger effect than Wavefront Healing.

(6) It is imperative to break the modal degeneracy in the SRC (and also to avoid accidental resonances there), in order to avoid the Wavefront Harming effect.

(7) The DR tuning known as RSE is especially sensitive to high scattering losses in the FP arm cavities, because it is designed to greatly increase the stored FP-arm power; and it is also especially sensitive to the Wavefront Harming effect, since a degenerate SEC cavity tuned to RSE would resonantly amplify all non-$TEM_{00}$ loss modes at the interferometer exit port.

In conclusion, we do find that Dual Recycling can be an effective system for obtaining sharp GW-sensitivity peaks, and for reducing the losses at the exit port of the interferometer by a significant factor. Its overall benefits, however, are severely constrained by the limitations of the optics, especially by fine-scale mirror surface roughness (as well as by mirror internal thermal noise). Rather than making the system more tolerant of optical deformations, as has been suggested [e.g., 9] -- and which would be true if one solely considered mirror tilt and curvature errors -- we argue here that Dual Recycling provides a renewed impetus for producing the highest quality of mirrors that can reasonably be achieved, in order to get the most possible benefit from this powerful configuration for advanced GW-interferometers.

# Acknowledgments

I would like to thank David Shoemaker and Yaron Hefetz for their advice and support during much of this research; and Hughes-Danbury and Corning for their contribution of mirror map data. The development of the numerical simulation program was principally supported by NSF Cooperative Agreement PHY-9210038.

# **References**[9]

[1] A. Abramovici *et al.*, Science **256**, 325 (1992).

[2] A. Giazotto, in *First Edoardo Amaldi Conference on Gravitational Wave Experiments*, edited by E. Coccia, G. Pizella, and F. Ronga (World Scientific, Singapore, 1995), p. 86.

[3] K. Danzmann, *ibid.,* p. 100.

[4] K. Tsubono, *ibid.*, p. 112.

[5] D. G. Blair, J. Munch, D. E. McClelland, and R. J. Sandeman, Australian Consortium for Interferometric Gravitational Astronomy, ARC Project, 1997 (unpublished).

[6] K. Thorne, LIGO Tech. Doc. P000024-00-R, 2000 (unpublished).

[7] B. J. Meers, Phys. Rev. D **38**, No. 8, 2317 (1988).

[8] B. J. Meers, Phys. Lett. A **142**, No. 8-9, 465 (1989).

[9] B. J. Meers and K. A. Strain, Phys. Rev. D **43**, No. 10, 3117 (1991).

[10] A. J. Tridgell, D. E. McClelland, and C. M. Savage, in *Gravitational Astronomy: Instrument Design and Astrophysical Prospects*, edited by D. E. McClelland and H.-A. Bachor (World Scientific, Singapore, 1991), p. 222.

[11] A. J. Tridgell, D. E. McClelland, C. M. Savage, and B. J. Meers, in *Proceedings of the Sixth Marcel Grossman Meeting on General Relativity*, edited by H. Sato and T. Nakamura (World Scientific, Singapore, 1992), p. 218.

[12] D. E. McClelland, C. M. Savage, A. J. Tridgell, and R. Mavaddat, Phys. Rev. D **48**, No. 12, 5475 (1993).

[13] R. Mavaddat, D. E. McClelland, P. Hello, and J.-Y. Vinet, J. Optics (Paris) **26**, No. 4, 145 (1995).

---

9. Referenced LIGO Technical Documents are available at "http://admdbsrv.ligo.caltech.edu/dcc/".

[14] D. E. McClelland, Aust. J. Phys. **48**, 953 (1995).

[15] B. Petrovichev, M. Gray, and D. McClelland, Gen. Rel. Grav. **30**, No. 7, 1055 (1998).

[16] K. A. Strain and B. J. Meers, Phys. Rev. Lett. **66**, No. 11, 1391 (1991).

[17] G. Heinzel *et al.*, Phys. Lett. A **217**, 305 (1996).

[18] G. Heinzel *et al.*, Phys. Rev. Lett. **81**, No. 25, 5493 (1998).

[19] M. B. Gray, A. J. Stevenson, H.-A. Bachor, and D. E. McClelland, Appl. Opt. **37**, No. 25, 5886 (1998).

[20] D. A. Shaddock, M. B. Gray, and D. E. McClelland, Appl. Opt. **37**, No. 34, 7995 (1998).

[21] A. Freise *et al.*, Phys. Lett. A **277**, 135 (2000).

[22] D. E. McClelland *et al.*, Class. Quant. Grav. **18**, 4121 (2001).

[23] T. Delker, "Summary of Results for Florida's Dual-Recycled Cavity-Enhanced Michelson Tabletop Prototype", www.phys.ufl.edu/LIGO/LIGO/DOCS/DRreport.pdf, Feb. 2001 (unpublished).

[24] O. Miyakawa, K. Somiya, G. Heinzel, and S. Kawamura, Class. Quant. Grav. **19**, 1555 (2002).

[25] G. de Vine, D. A. Shaddock, and D. E. McClelland, Class. Quant. Grav. **19**, 1561 (2002).

[26] B. Willke *et al.*, Class. Quant. Grav. **19**, 1377 (2002).

[27] G. Heinzel *et al.*, Class. Quant. Grav. **19**, 1547 (2002).

[28] W. Winkler, K. Danzmann, A. Rüdiger, and R. Schilling, in *Proceedings of the Sixth Marcel Grossman Meeting on General Relativity*, edited by H. Sato and T. Nakamura (World Scientific, Singapore, 1992), p. 176.

[29] B. Bochner, Ph.D. thesis, Massachusetts Institute of Technology, 1998; B. Bochner, LIGO Tech. Doc. P980004-00-R, 1998 (unpublished).

[30] B. Bochner and Y. Hefetz, "A grid-based simulation program for gravitational wave interferometers with realistically imperfect optics", submitted to Phys. Rev. D (2002).

[31] S. Whitcomb *et al.*, in *Proceedings of the TAMA Workshop on Gravitational Wave Detection*, edited by K. Tsubono (Universal Academic Press, Tokyo, 1997).

[32] J. B. Camp *et al.*, in *Laser Induced Damage in Optical Materials*, SPIE Proceedings Vol. 4679, 1 (2002).

[33] K. Ganezer, LIGO Tech. Doc. G010119-00-Z, 2001 (unpublished).

[34] B. Kells and J. Camp, LIGO Tech. Doc. T970097-01-D, 1997 (unpublished).

[35] D. E. McClelland *et al.*, Opt. Lett. **24**, No. 15, 1014 (1999).

[36] T. Tomaru *et al.*, Appl. Opt.-OT **41**, No. 28, 5913 (2002).

[37] T. Tomaru *et al.*, Class. Quant. Grav. **19**, No. 7, 2045 (2002).

[38] J. Mizuno *et al.*, Phys. Lett. A **175**, No. 5, 273 (1993).

[39] E. Gustafson, D. Shoemaker, K. Strain, and R. Weiss, LIGO Tech. Doc. T990080-00-D, 1999 (unpublished).

[40] G. Billingsley, LIGO Tech. Doc. E000041-A-D, 2001 (unpublished).

[40.5] P. Fritschel, in *Gravitational-Wave Detection*, SPIE Proceedings Vol. 4856, edited by P. Saulson and M. Cruise (2003).

[41] A. Weinstein, Class. Quant. Grav. **19**, 1575 (2002).

[42] R. W. P. Drever, in *Gravitational Radiation*, NATO Advanced Physics Institute, Les Houches, edited by N. Deruelle and T. Piran (North-Holland, Amsterdam, 1983), p. 321.

[43] R. E. Vogt *et al.*, "Proposal to the National Science Foundation: The Construction, Operation, and Supporting Research and Development of a Laser Interferometer Gravitational-Wave Observatory", 1989 (unpublished).

[44] A. Krolak, J. A. Lobo, B. J. Meers, Phys. Rev. D **43**, No. 8, 2470 (1991).

[45] J.-Y. Vinet, B. J. Meers, C. N. Man, and A. Brillet, Phys. Rev. D **38**, No. 2, 433 (1988).

[46] A. E. Siegman, *Lasers* (University Science Books, California, 1986).

[47] M. C. Weisskopf, Astrophys. Lett. Comm. **26**, 1 (1987).

[48] Y. Hefetz, personal communications.

[49] W. H. Press, S. A. Teukolsky, W. T. Vetterling, and B. P. Flannery, *Numerical Recipes in Fortran: The Art of Scientific Computing* (Cambridge University Press, Cambridge, and Numerical Recipes Software, 1992).

[50] P. Saha, J. Opt. Soc. Am. A **14**, No. 9, 2195 (1997).

[51] K. A. Strain, "LIGO II Configuration options", www.phys.ufl.edu/LIGO/LIGO/DOCS/config_1.pdf, Apr. 2000 (unpublished).

[52] J. Mason and P. Willems, in *Gravitational Waves: Third Edoardo Amaldi Conference*, edited by S. Meshkov (American Institute of Physics, 2000), p. 203.

[53] M. B. Gray, D. A. Shaddock, and D. E. McClelland, in *Gravitational Waves: Third Edoardo Amaldi Conference*, edited by S. Meshkov (American Institute of Physics, 2000), p. 193.

[54] E. E. Flanagan and S. A. Hughes, Phys. Rev. D **57**, No. 8, 4535 (1998).

[55] K. C. B. New *et al.*, Astrophys. J. **450**, 757 (1995).

[56] K. S. Thorne, in *Proceedings of the Snowmass 95 Summer Study on Particle and Nuclear Astrophysics and Cosmology*, edited by E. W. Kolb and R. Peccei (World Scientific, Singapore, 1995), p. 160.

# Table Captions

**TABLE I.** Optical parameters used for the simulation runs presented in this paper. (Notes in bold refer to "Config. II" Long-SRC runs; cf. Sec. 5.2, Fig's 10, 11.) The labeling of the optical elements are as depicted in Fig. 1.

**Table I**

**Brett Bochner**

**Simulating a Dual-Recyc. GW Interferometer w/Realistically Imperfect Optics**

| Quantity | Value(s) |
|---|---|
| Carrier Laser Wavelength | 1.064 μm (Nd:YAG light) |
| Carrier Laser Power | Results normalized to 1 Watt |
| Cavity Lengths | $L_1 = 5$ m <br> $L_2 = L_3 = 4.19$ m <br> $L_4 = L_5 = 4.0$ km <br> $L_6 = 5$ m (*or **2004.19 m***) |
| Mirror Curvature Radii | $R_{curv,\,1} = 9.99$ km <br> $R_{curv,\,2} = R_{curv,\,3} = 14.6$ km <br> $R_{curv,\,4} = R_{curv,\,5} = 7.4$ km <br> $R_{curv,\,6} = 14.1$ km (*or **15.4 km***) |
| Mirror Intensity Reflectivities (Reflective Side) | $R_1 \sim .94 - .985$ (*Optimized For Max. PRC Power*) <br> $R_2 = R_3 = .97$ <br> $R_4 = R_5 = .99994$ <br> $R_{bs} = .49992$ <br> $R_6 \sim 0 - .99$ (*Varied*) |
| Mirror Intensity Reflectivities (Anti-reflective Side) | $R_1, R_6$ : Same as Ref-Sides <br> $R_2 = R_3 = .968817$ <br> $R_{bs} = .49971$ |
| Mirror Intensity Transmissions (Both Sides) (Pure Loss $\equiv 1 - R - T$) | $T_1 = 1 - R_{1,\,Optim.} - 50$ ppm Loss <br> $T_2 = T_3 = .02995$ <br> $T_{bs} = .50003$ <br> $T_6 = 1 - R_6 - 50$ ppm Loss |
| Beam Waist Diameter | 7.02 cm |
| Mirror Aperture Diameters | 24 cm (Circular Mirrors) <br> (*Or **23.5 cm**, for $R_6$*) <br> 24.4 tilted 45° (Beamsplitter) <br> (*Or All Varied, ~ 24 - 70 cm*) |
| Mirror Thicknesses (Perpendicular to Surface) | Beamsplitter = 4 cm <br> All Others = 10 cm |
| Substrate Refraction Index | n = 1.44963 |

# Figure Captions

**FIG. 1.** Schematic diagram of the core optical configuration of a LIGO interferometer, incorporating (in dotted lines) a Signal Recycling mirror for DR. (Not drawn to scale.)

**FIG. 2.** The GW-signal amplitude (unnormalized, using $h = 1$ for convenience), proportional to the summed amplitudes of the "plus" and "minus" GW-induced sideband fields emerging from the interferometer exit port, is plotted versus GW-frequency ($f_{GW}$) for curves representing different Signal Recycling Mirror reflectivities. The $R_{dual}$ values are (in order of increasing peak sharpness): 0.0, 0.1, 0.3, 0.5, 0.7, 0.9, 0.99. The DR system is tuned to $\nu_{GW}^{optim.} = 500$ Hz for all curves.

**FIG. 3.** The (unnormalized) GW-signal amplitudes, plotted versus $f_{GW}$, for different SRC-tuning optimization frequencies. From left to right, the curves are for $\nu_{GW}^{optim.}$ equal to (in Hz): 0, 200, 500, 750, and 900. $R_{dual} = 0.9$ for all curves.

**FIG. 4.** Power lost through the exit port (normalized to 1 Watt of interferometer excitation laser power), plotted versus $R_{dual}$, for Broadband DR.

**FIG. 5.** The (unnormalized) GW-signal amplitudes (TEM$_{00}$ mode only), plotted versus $f_{GW}$, for $\nu_{GW}^{optim.} = 200$ Hz. Each box of plots uses a distinct value of $R_{dual}$, from 0.0 to 0.99. The plots compare the theoretically-calculated response curves (solid lines), with results from runs of our simulation program (dotted curves). Each set of three dotted curves, from highest to lowest, represents runs with: (i) "perfect" mirror surfaces and substrates; (ii) $\lambda/1800$ surfaces (and deformed substrates); (iii) $\lambda/800$ surfaces (and deformed substrates).

**FIG. 6.** The (unnormalized) GW-signal amplitudes (TEM$_{00}$ mode only), plotted versus RMS mirror surface deformation amplitude, for different values of $R_{dual}$. Each data point is a *ratio* of the GW-signal amplitude (at $f_{GW} = \nu_{GW}^{optim.} = 200$ Hz) for a particular deformed-mirrors run, divided by the GW-signal amplitude for the perfect-mirrors run, for that *same* value of $R_{dual}$. Each solid line connects the results for a particular value of $R_{dual}$, from 0.0 to 0.99.

**FIG. 7.** Interferometer exit-port power losses (all modes), plotted versus RMS mirror surface deformation amplitude, for $R_{dual}$ values from 0.0 to 0.99. $\nu_{GW}^{optim.} = 200$ Hz for all curves.

**FIG. 8.** Resonant power buildup in the inline Fabry-Perot arm cavity, plotted versus $R_{\text{dual}}$, for different tilts of the arm cavity back mirrors ($\nu_{\text{GW}}^{\text{optim.}} = 200$ Hz for all curves). The effects of Wavefront Healing are evident.

**FIG. 9.** Resonant power buildup in the inline FP arm cavity, plotted versus the aperture radii of all interferometer mirrors ($\nu_{\text{GW}}^{\text{optim.}} = 200$ Hz for all curves). Results for 4 cases are shown: "perfect" and "deformed" mirror maps, with and without a Signal Recycling Mirror.

**FIG. 10.** Interferometer exit-port power losses (all modes), plotted versus $R_{\text{dual}}$, for different configurations of the Signal Recycling Cavity. (All runs use $\lambda/800$ mirror surfaces and deformed substrates; $\nu_{\text{GW}}^{\text{optim.}} = 1000$ Hz for all curves). The effects of SRC degeneracy and degeneracy-breaking (with a long SRC cavity) are evident.

**FIG. 11.** Resonant power buildup in the inline FP arm cavity (all modes), plotted versus $R_{\text{dual}}$, for different SRC configurations. (All runs use $\lambda/800$ mirror surfaces and deformed substrates; $\nu_{\text{GW}}^{\text{optim.}} = 1000$ Hz for all curves). The effects of "Wavefront Healing" (for nondegenerate SRC) and "Wavefront Harming" (for degenerate SRC) are evident.

**Figure 1**
**Brett Bochner**
**Simulating a Dual-Recyc. GW Interferometer w/Realistically Imperfect Optics**

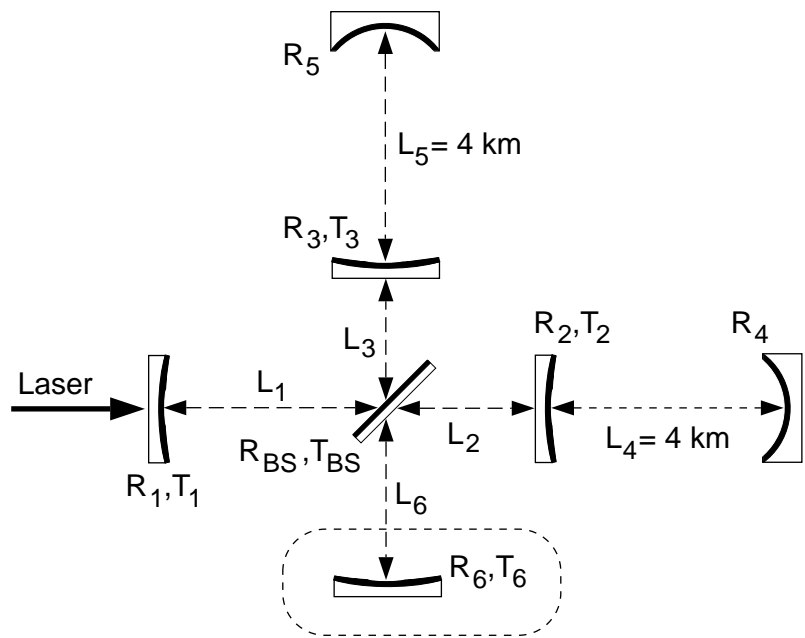

**Figure 2**
**Brett Bochner**
**Simulating a Dual-Recyc. GW Interferometer w/Realistically Imperfect Optics**

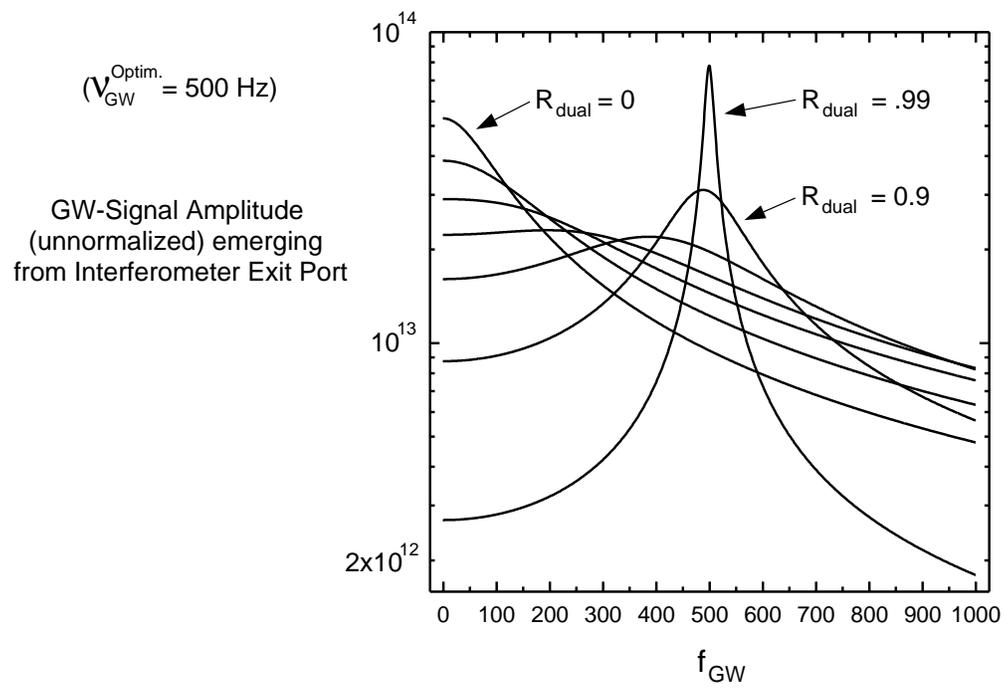

**Figure 3**
**Brett Bochner**
**Simulating a Dual-Recyc. GW Interferometer w/Realistically Imperfect Optics**

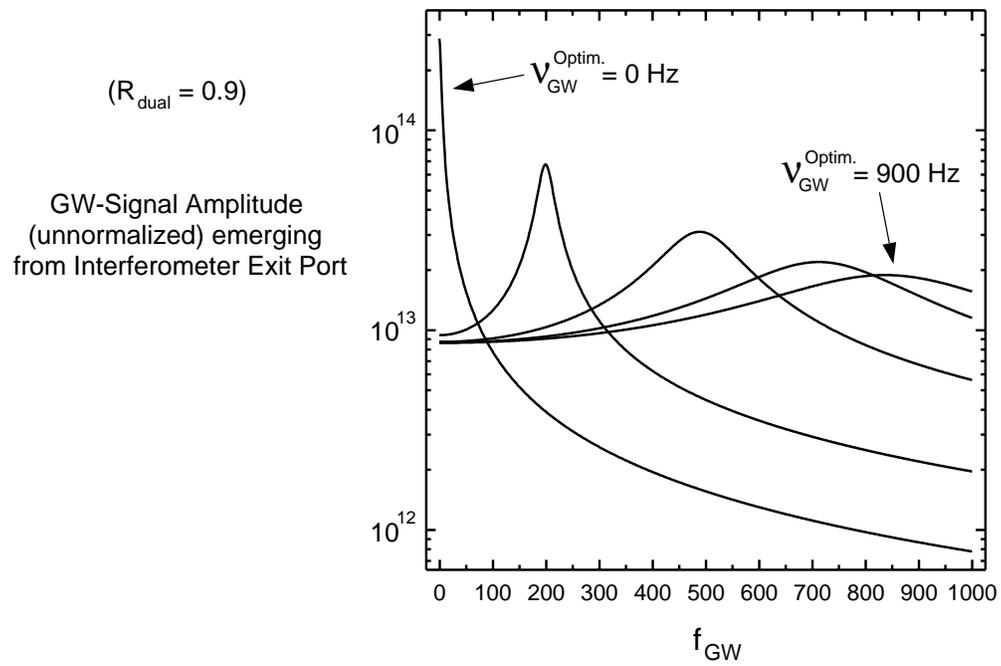

**Figure 4**
**Brett Bochner**
**Simulating a Dual-Recyc. GW Interferometer w/Realistically Imperfect Optics**

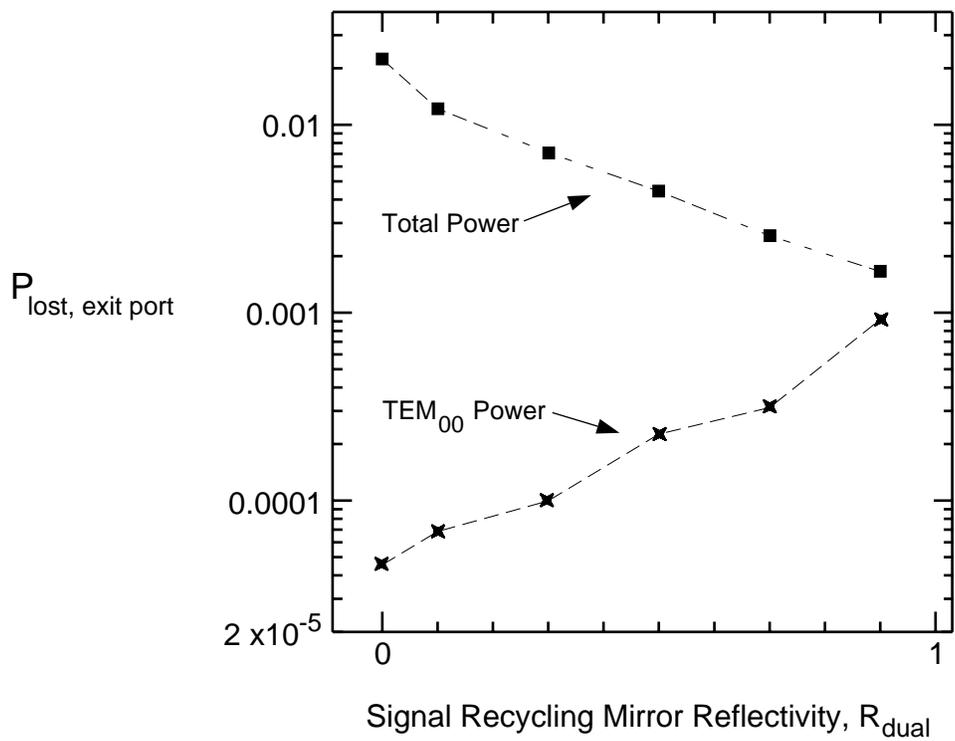

**Figure 5
Brett Bochner
Simulating a Dual-Recyc. GW Interferometer w/Realistically Imperfect Optics**

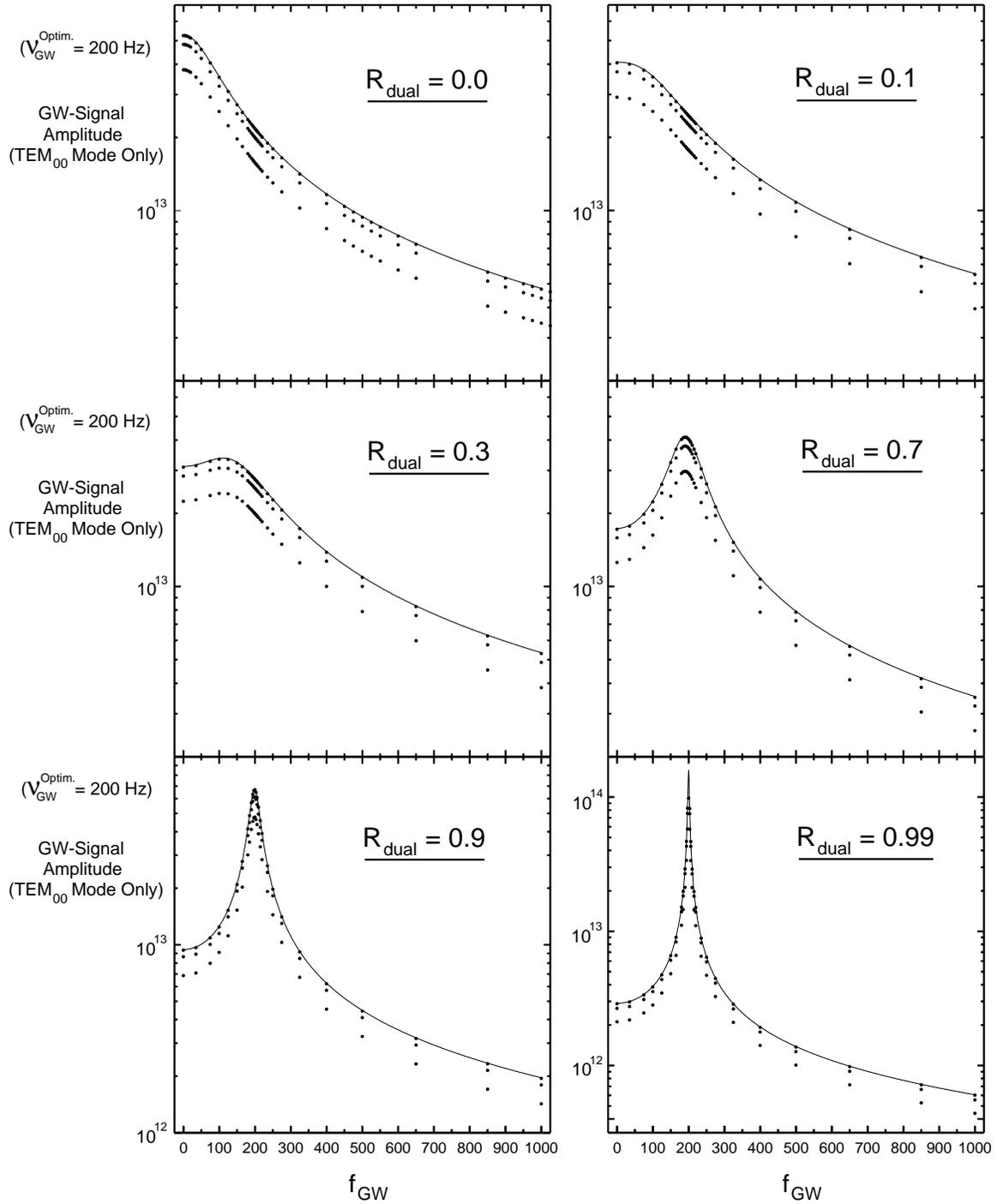

**Figure 6**
**Brett Bochner**
**Simulating a Dual-Recyc. GW Interferometer w/Realistically Imperfect Optics**

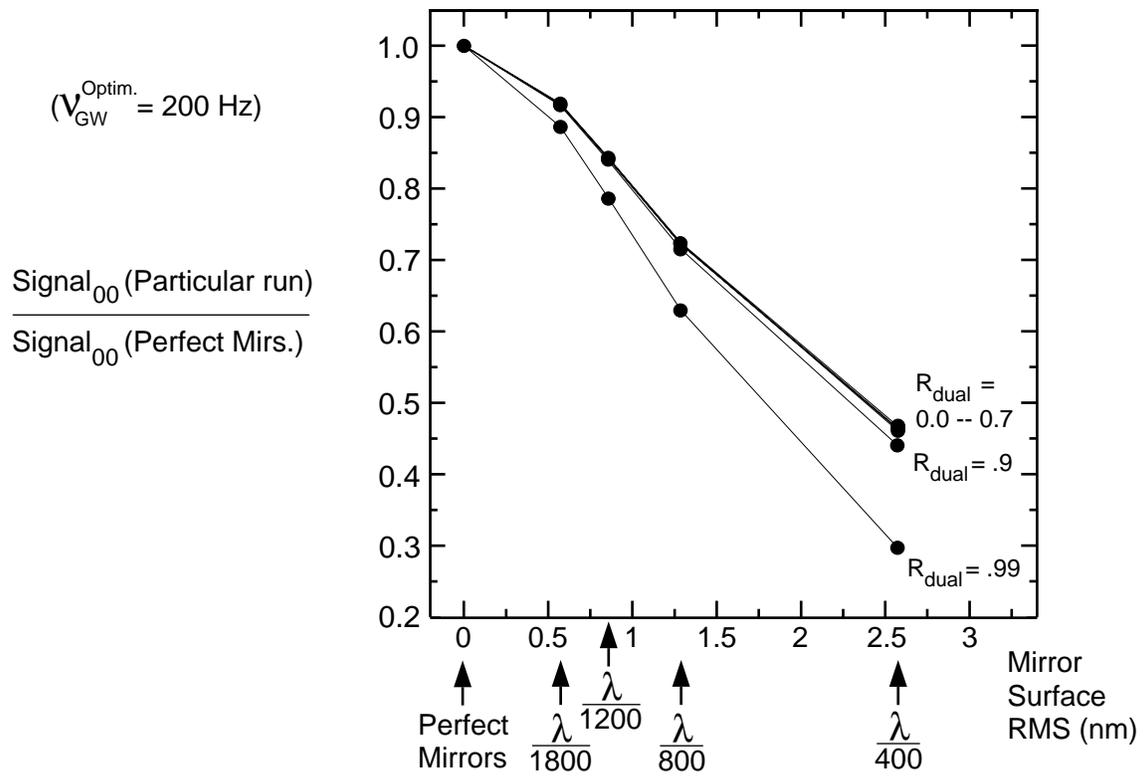

**Figure 7**
**Brett Bochner**
**Simulating a Dual-Recyc. GW Interferometer w/Realistically Imperfect Optics**

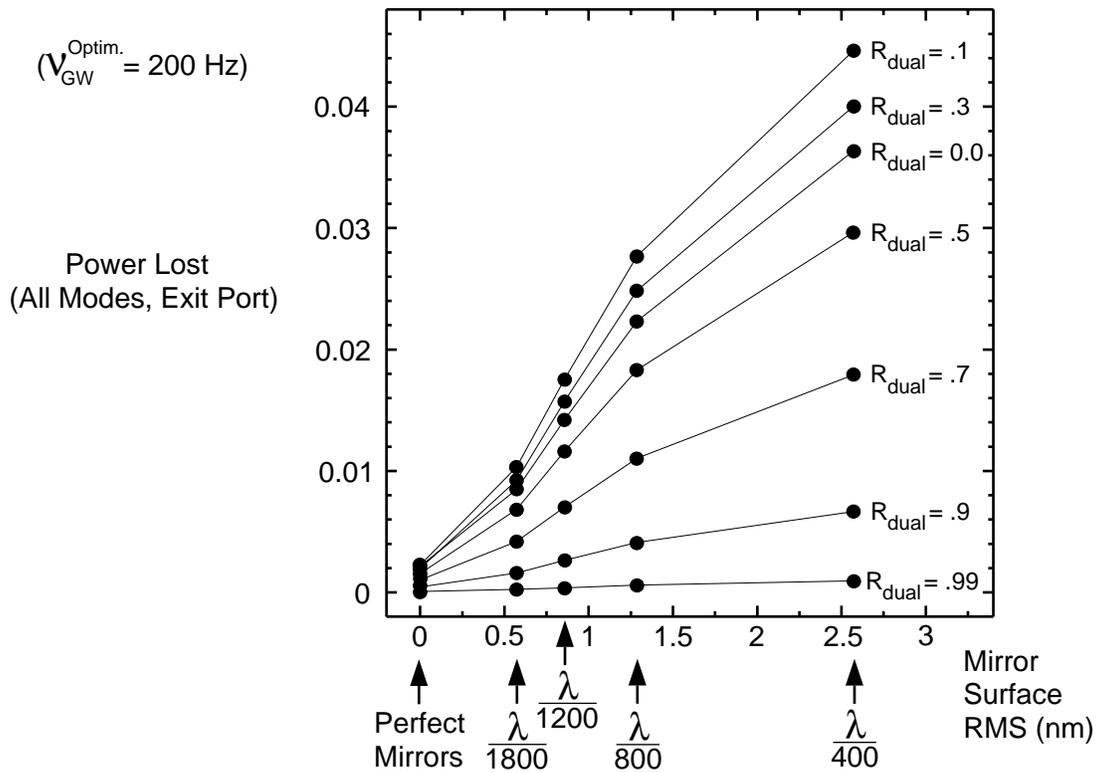

**Figure 8**
**Brett Bochner**
**Simulating a Dual-Recyc. GW Interferometer w/Realistically Imperfect Optics**

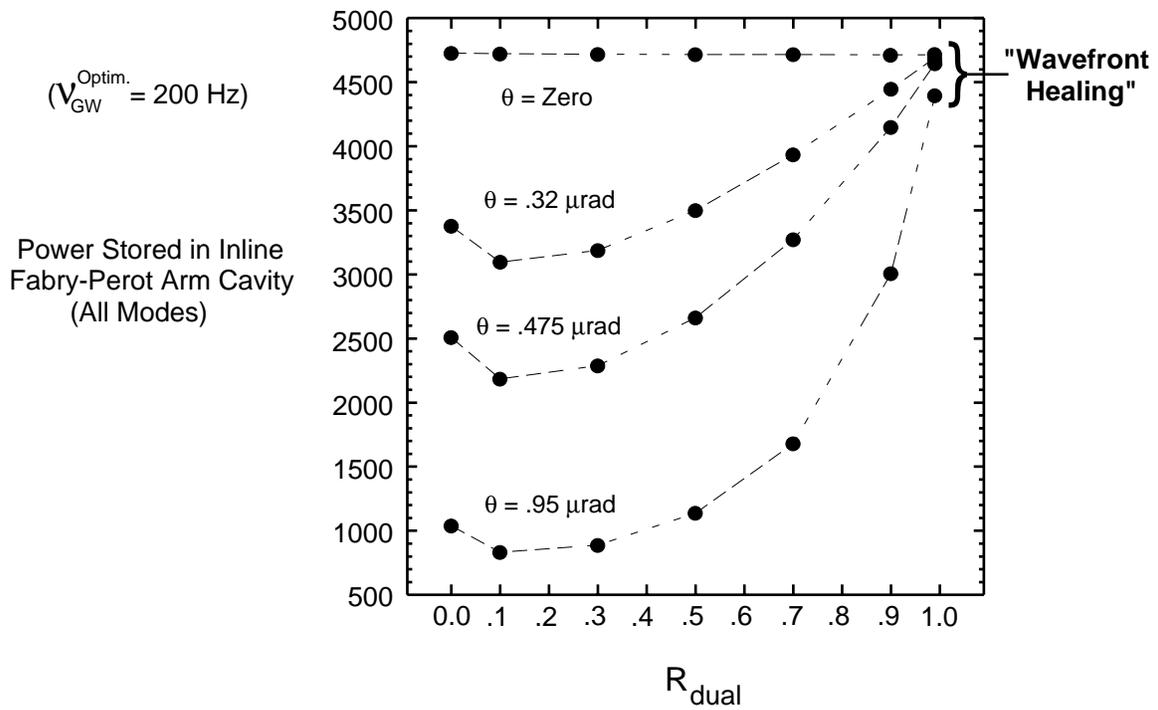

**Figure 9**
**Brett Bochner**
**Simulating a Dual-Recyc. GW Interferometer w/Realistically Imperfect Optics**

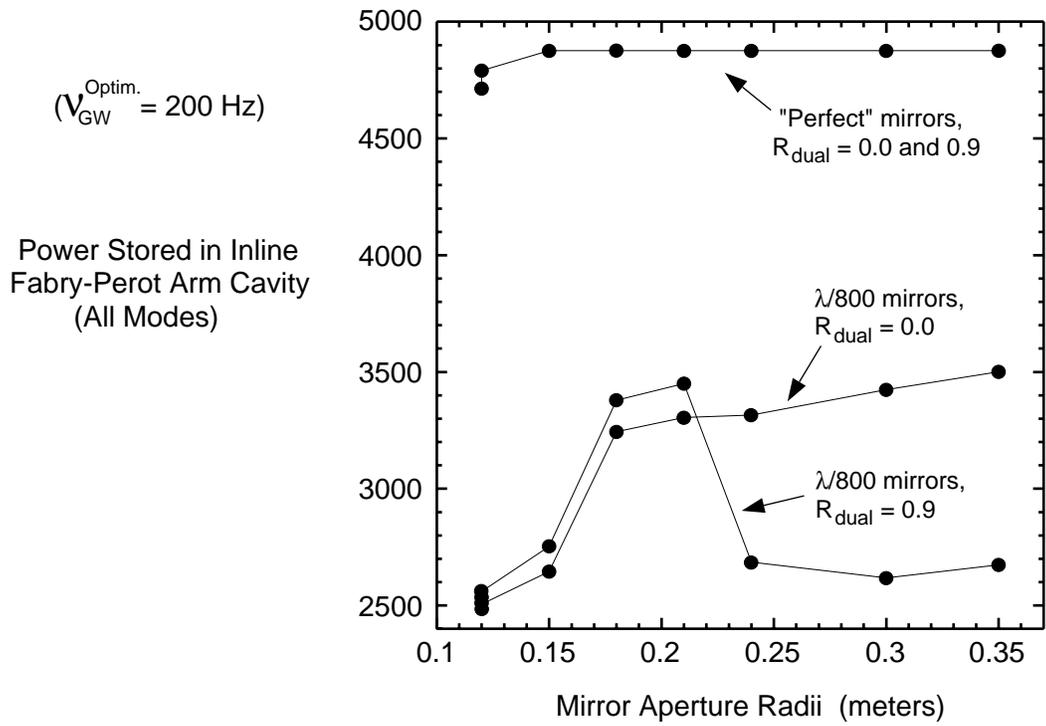

**Figure 10**
**Brett Bochner**
**Simulating a Dual-Recyc. GW Interferometer w/Realistically Imperfect Optics**

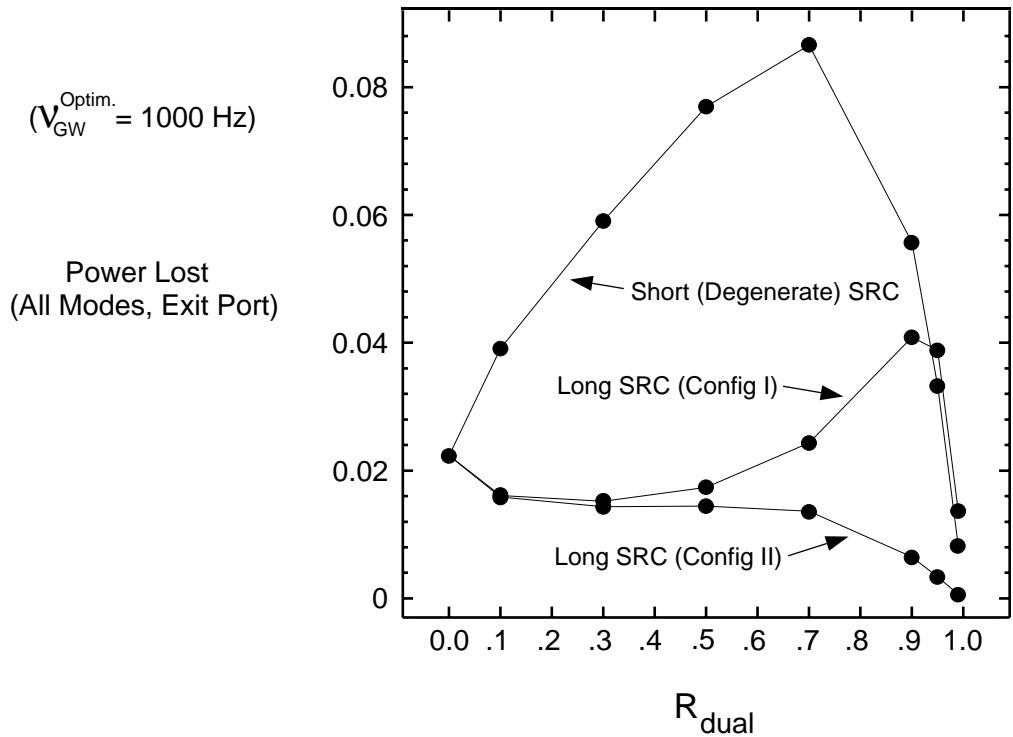

**Figure 11**
**Brett Bochner**
**Simulating a Dual-Recyc. GW Interferometer w/Realistically Imperfect Optics**

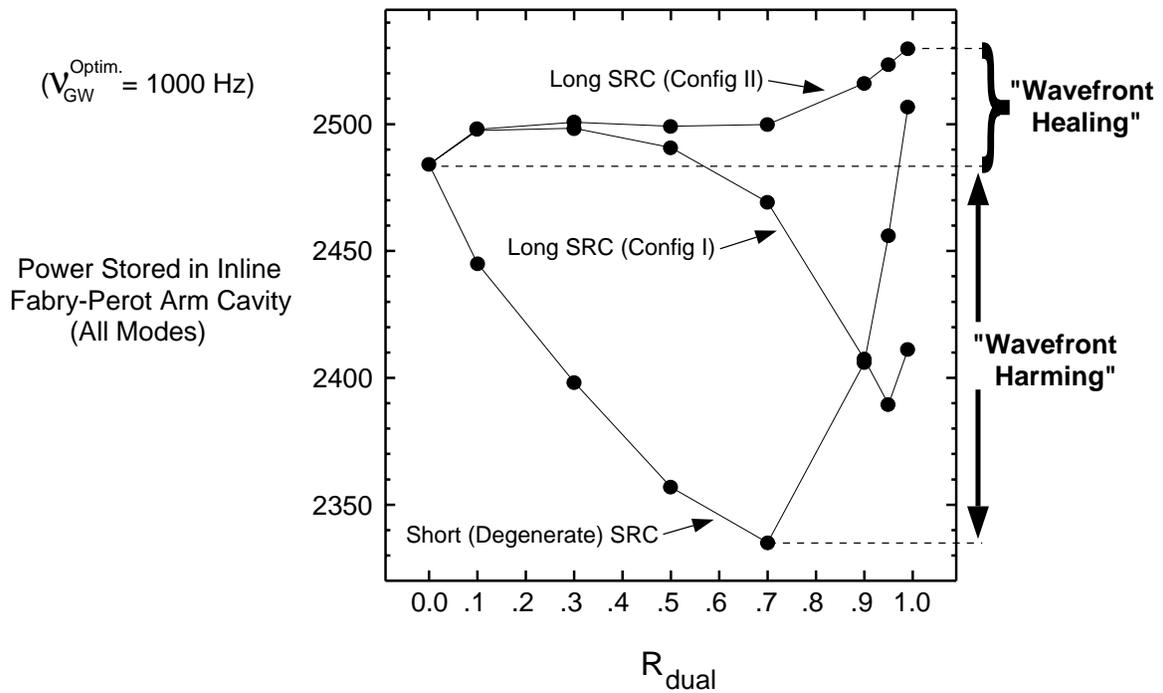